\documentclass[aps,prd,preprint,groupedaddress]{revtex4}
\usepackage{epsf,epsfig}
\newcommand{\Exp}[1]{{\mathrm{e}}^{#1}}
\newcommand{\be}{\begin{equation}}
\newcommand{\ee}{\end{equation}}
\newcommand{\bea}{\begin{eqnarray}}
\newcommand{\eea}{\end{eqnarray}}
\newcommand{\LL}{{\mathcal L}}
\newcommand{\pa}{\partial}
\newcommand{\pam}{{\partial_\mu}}

\newcommand{\ba}{\begin{array}}
\newcommand{\ea}{\end{array}}
\newcommand{\lp}{\left(}
\newcommand{\rp}{\right)}

\newcommand{\g}{\gamma}

\newcommand{\om}{\omega}

\newcommand{\ep}{\epsilon}

\begin{document}

                
\title{Interactions of Q-balls and matter.}


\author{Stephen S. Clark}
\affiliation{Institut de th\'eorie des ph\'enom\` enes physiques, Ecole Polytechnique F\'ed\'erale de Lausanne, CH-1015 Lausanne, Switzerland.}


\date{\today}

\begin{abstract}
We know from previous work \cite{clark} that non topological solitons, Q balls, evaporate into fermions. All the constructions we used to find evaporation rate
were dased on the fact that no fermions would move towards the Q ball. All these constructions left an opened question that is : what happens when a fermion 
interacts with a Q ball. We shall answer this question in this work by using the constructions done to compute evaporation rates. We shall also obtain a new
approach to compute evaporation rates.
\end{abstract}


\maketitle
\section{Introduction}
A scalar field theory with an unbroken continuous global symmetry admits a remarkable class of solutions, non-topological solitons or
Q-Balls. These solutions are spherically symmetric non-dissipative solutions to the classical field equations \cite{Qballscoleman,Qballscohen,Qballsnew,clark}.
These solutions are constructed using the fact that they are absolute minima of the energy, this property enshures the stability of Q-balls 
versus decay into scalars. The addition of a coupling to fermions will result in Q-balls becoming unstable versus decay into fermions \cite{Qballscohen,Qballsnew,clark}.
The result of this instability is Q-ball evaporation.

Q-ball evaporation admits an absolute upper bound resulting from the fact that all the energy ranges are finate. 
The fact that energy ranges are finate comes from the mixing of positive and negative frequency terms. We could ask, what would happen if the fermions
interacting with the Q-ball have an energy lying outside the evaporation range? This is the question we propose to answer in this work. We know from 
previous work \cite{clark} that we can use two methods to obtain a solution to the equations of motion describing a fermion interacting with a Q-ball. Our
starting point will be the solutions expressed in terms of partial waves, that we shall study outside the evaporation range. This partial wave superposition
has the reflection and transmission amplitudes as expansion coefficients. So we shall study the diffusion of a fermion on a Q ball.
We shall then apply the same calculations to the total energy spectrum to see to the interference of both phemomenoms (evaporation and diffusion).

In fact we expect the difficulties to appear when the energy of the interacting fermion lies inside the evaporation range for this reason we shall start with 
fermions having their energy bigger than the evaporation upper bound. In this situation expect that the only wave having a non-trivial behaviour would be
one going through the Q ball. The second regime where the energy of the interacting fermion lies inside the evaporation range will lead to some new results.

The work will be organised as follows we shall first give a brief overview of the method to construct the solutions. We shell then study these solutions outside
the evaporation range and then we shall try to look at the total energy range. 
\section{Preliminaries} 
The details of all the following calculations can be found in previous work \cite{Qballscohen,Qballsnew,clark} and we shall only give here a brief
overview of the results and methods used. The major idea to solve this problem is considering a Q ball localized in space, so the fermionic field
equations are free ones outside the zone where we have Yukawa coupling (interaction of Q ball and fermions). For the two cases we shall consider,
massless and massive fermions, we shall need to match the solutions outside to the solutions inside the Q ball. The matching coefficients will give the reflection
and transmission amplitudes we shall use in the problem.
\subsection{Solutions to the equations of motion, massless case}
Writing down the Lagrangian of a massless fermion having a Yukawa interaction with a scalar field gives in one spatial dimension, 
\begin{eqnarray}
\LL_{ferm.}=i\bar{\psi}\sigma^\mu\pam\psi+(g\phi\bar{\psi}^C\psi+h.c),
\end{eqnarray}
Equations of motion for the two components of the $\Psi$ field are : 
\begin{eqnarray}
\ba{c}(i\pa_0+i\pa_z)\psi_1-g\phi\psi_2^{\star}=0, \\
(i\pa_0-i\pa_z)\psi_2^{\star}-g\phi^\star\psi_1=0. \ea
\end{eqnarray}
and $\phi=\phi_0\Exp{-i\om_0t}$ in the zone from $-l$ to $+l$ and zero everywhere else. 
Using the ansatz :
\begin{eqnarray}
\lp \ba{c} \psi_1 \\ \psi_2^{\star} \ea \rp=\lp \ba{cc} \Exp{-i\frac{\om_0}{2}t} & 0\\
0 & \Exp{i\frac{\om_0}{2}t} \ea \rp \lp\ba{c} A  \\
B\ea\rp\Exp{-i\ep t+i(k+\frac{\om_0}{2})},\label{ansatz}
\end{eqnarray}
the equations of motion are reduced to the following $2\times2$ linear system
\begin{eqnarray}
\lp\ba{cc} k-\ep & M \\ M & -(k+\ep)\ea\rp\lp\ba{c} A \\ B \ea\rp=0 . \nonumber
\end{eqnarray}
The determinant of the system gives $k=\pm\sqrt{\ep^2-M^2}\equiv \pm k_\ep$.
Solving for the two cases  $k=+k_\ep$ and $k=-k_\ep$, we obtain the solution inside the Q-Ball:
\begin{eqnarray}
\Psi_Q=\lp\ba{c}\psi_1\\ \psi_2^\star\ea\rp=
A\lp\ba{c} 1 \\ \frac{k_\ep+\ep}{M} \ea\rp\Exp{-ik_\ep z} + B\lp\ba{c} \frac{k_\ep+\ep}{M} \\ 1 \ea\rp\Exp{ik_\ep z},
\label{inside}
\end{eqnarray}
where $M=g\phi_0$, $g$ is the coupling constant and $\phi_0$ the value of the scalar field. 
The second part is the  solution when $\phi_0=0$ (outside the Q-Ball) it is,
\begin{eqnarray}
\Psi=\lp\ba{c} \psi_1 \\ \psi_2^\star \ea \rp=
\Exp{-i\ep t}\lp\ba{c} C_1^{L,R}\Exp{i\ep z} \\ C_2^{L,R}\Exp{-i\ep z}\ea\rp, \label{coeflibre}
\end{eqnarray}
where superscripts $L,R$ indicate the left and right side of the Q-Ball. In order to solve Dirac's equation everywhere
the solution needs to be continuous in space. Space continuity gives at $z=-l$ :
\begin{eqnarray}
C_1^L=A\Exp{i(k_\ep+\ep)l}+B\alpha_\ep\Exp{-i(k_\ep-\ep)l}, \nonumber \\
C_2^L=A\alpha_\ep\Exp{i(k_\ep-\ep)l}+B\Exp{-i(k_\ep+\ep)l}, \nonumber
\end{eqnarray}
and at $z=+l$,
\begin{eqnarray}
C_1^R=A\Exp{-i(k_\ep+\ep)l}+B\alpha_\ep\Exp{+i(k_\ep-\ep)l}, \nonumber\\
C_2^R=A\alpha_\ep\Exp{-i(k_\ep-\ep)l}+B\Exp{+i(k_\ep+\ep)l}. \nonumber
\end{eqnarray}
These two relations will give the matrix linking the coefficients on the left to those on the right.
\subsection{Solution to the motion equations, massive case.}
Using the same Lagrangian as for the massless case and adding a Dirac mass coupling for massive fermions, gives the fermionic
Lagrangian : 
\begin{eqnarray}
\LL=\bar{\psi}i\g^\mu\pam\psi+g(\bar{\psi}^C\psi\phi+h.c.)+M_D(x)\bar{\psi}\psi.
\end{eqnarray}
The equations of motion using two component $\psi$-field, 
in $1\oplus1$-dimensions are
\begin{eqnarray}
\ba{c}(i\pa_t+i\pa_z)\psi_1-M\Exp{-i\frac{\om_0}{2}t}\psi_2^\star+M_D\psi_2=0, \\
(i\pa_t-i\pa_z)\psi_2+M\Exp{-i\frac{\om_0}{2}t}\psi_1^\star+M_D\psi_1=0. \ea \label{mass1}
\end{eqnarray}
These equations can be solved if we use fields with four degrees of freedom. Solving this system will give us the
solution inside the Q-Ball, which is the first step. To solve these equations of motion we use 
the following ansatz
\begin{eqnarray}
\ba{c}\psi_1=f_1(z)\Exp{i(\ep-\frac{\om_0}{2})t}+f_2(z)\Exp{-i(\ep+\frac{\om_0}{2})t}, \\
\psi_2=g_1(z)\Exp{i(\ep-\frac{\om_0}{2})t}+g_2(z)\Exp{-i(\ep+\frac{\om_0}{2})t}. \ea \label{mass2}
\end{eqnarray}
Due to the separation of time components this ansatz will lead to four equations.
Taking,
\begin{eqnarray}
f_1(z)=A\Exp{ipz}, f_2^\star(z)=B\Exp{ipz}, g_1(z)=C\Exp{ipz} g_2^\star(z)=D\Exp{ipz}. 
\end{eqnarray}
and after some
re-arrangement of the equations we obtain :
\begin{eqnarray}
\ba{c}-\ep_-f_1-g_2^\star+M_Dg_1=pf_1, \\
\ep_-g_1-f_2^\star-M_Df_1=pg_1, \\
-\ep_+f_2^\star+g_1-M_Dg_2^\star=pf_2^\star, \\
\ep_+g_2^\star+f_1+M_Df_2^\star=pg_2^\star,\ea
\end{eqnarray}
where $\ep_-=\ep-\frac{\om_0}{2}$ and $\ep_+=\ep+\frac{\om_0}{2}$.
This arrangement has the advantage that we can now write the $\psi$-field in terms of four component spinors as :
\begin{eqnarray}
\Psi=\lp\ba{c}\lp\ba{c}f_1\\g_1\ea\rp\\
\lp\ba{c}f_2^\star\\g_2^\star\ea\rp\ea\rp.
\end{eqnarray}
The equations of motion become in matrix form,
\begin{eqnarray}
\lp\ba{cccc} -\ep_- & M_D & 0 & -1 \\ -M_D & \ep_-& -1 & 0 \\
0& 1 & -\ep_+& -M_D \\ 1 & 0 & M_D & \ep_+\ea\rp\lp\ba{c} C_1 \\C_2 \\ C_3 \\ C_4 \ea\rp= 
Mp\lp\ba{c} C_1 \\ C_2 \\ C_3 \\ C_4 \ea\rp.\label{qballmatrix1}
\end{eqnarray}
All the parameters are normalised by $M$.
Inside the Q-ball the time dependent solution is :
\begin{eqnarray}
\Psi=\sum_{j=1}^{4}C_j\Exp{i(\ep-\om)t}v_{p_j}^{up}\Exp{i\bar{p}_jz}+C^\star_j\Exp{-i(\ep+\om)t}(v_{p_j}^{down}\Exp{i\bar{p}_jz})^\star,
\end{eqnarray}
where the $up$ superscript stands for the first two components of the eigenvectors, while the $down$ one indicates we
take the two last components of the eigenvectors of the inner motion matrix.
The time dependent solution outside the Q-ball is once more given by,
\begin{eqnarray}
\Psi=\sum_{j=1}^{4}(A_j,B_j)\Exp{i(\ep-\om)t}u_{p_j}^{up}\Exp{i\bar{p}_jz}+(A_j^\star,B_j^\star)\Exp{-i(\ep+\om)t}(u_{p_j}^{down}\Exp{i\bar{p}_jz})^\star,
\end{eqnarray}
where we this time use the eigenvectors of the outside motion matrix ($M_D=0$). The $A$'s $B$'s and $C$'s are expension coefficients. The
matching rules for these coefficients will give the reflection and transmission coefficients.
\section{Interactions of Q-balls and matter.} 
We know from previous work \cite{clark} that we can use two equivalent methods to solve the problem of particle creation from Q-balls. These two
methods allowed us to obtain the exact quantysed solutions describing fermions being produced by a Q-ball. The method we shall use now will consist in
using the total solutions outside the evaporation range $\ep\notin[-\frac{\om_0}{2}+M_D;\frac{\om_0}{2}+M_D]$. 
The solutions we constructed are superpositions of partial waves containing positive and
negative movers. 
\subsection{Interaction of Q ball and matter, massless case.}
We now study the diffusion processes on a Q ball. In the first part of the construction shall not consider any evaporation for Q balls. This simplification 
will allow us in the first place to build and study the diffusion of a fermion on a Q ball that we expect to be very similar as diffusion on a potential barrier.
The simplest aproach will be to consider all possible transmition trough
the Q ball, that is we might have a transmitted particle or anti-particle due to the energy shift (the $\frac{\om_0}{2}$ factor in the time
dependent solution).
\begin{figure}[h]
\begin{center}
\input{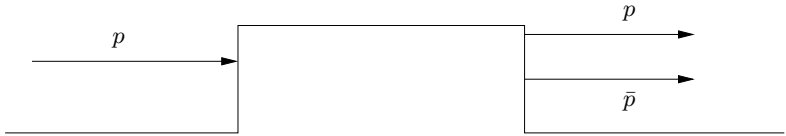}
\caption{General transmission through a Q ball.}
\end{center}
\end{figure}
For simplicity we shall consider the construction where there is no incoming anti-particle.

We start by using the solution outside the Q-Ball for the simple massles case we have:
\begin{eqnarray}
\Psi=\lp\ba{c} \psi_1 \\ \psi_2 \ea \rp=
\lp\ba{c} C_1^{L}\Exp{-i(\ep+\frac{\om_0}{2})t}\Exp{i(\ep+\frac{\om_0}{2})z} \\ 
(C_2^{L})^\star\Exp{i(\ep-\frac{\om_0}{2})t}\Exp{i(\ep-\frac{\om_0}{2})z}\ea\rp,
\end{eqnarray}
we have also a relation for coefficients on the right and on the left given by,
\begin{eqnarray}
\lp\ba{c} C_1^R \\ C_2^R \ea\rp=\frac{1}{1-\alpha^2}\lp\ba{cc} \Exp{-2ik_\ep l}-\alpha_\ep^2 \Exp{2ik_\ep l} & 
-\alpha_\ep\Exp{-2ik_\ep l}+\alpha_\ep\Exp{2ik_\ep l} \\\alpha_\ep\Exp{-2ik_\ep l}-\alpha_\ep\Exp{2ik_\ep l} & 
\Exp{2ik_\ep l}-\alpha_\ep^2 \Exp{-2ik_\ep l} \ea\rp\lp\ba{c} C_1^L \\ C_2^L \ea\rp. \nonumber
\end{eqnarray}
For the calculation of evaporation rate we considered that both components where the same particles what we do now is consider
both componets to be different particles (doing so we can construct the normal field containing both particles and anti-particles).
we shall consider as a first study :
\begin{eqnarray}
\ep+\frac{\om_0}{2}\geq0\rightarrow\ep\geq-\frac{\om_0}{2} \nonumber \\
\ep-\frac{\om_0}{2}\geq0\rightarrow\ep\geq\frac{\om_0}{2} \nonumber, 
\end{eqnarray}
it means we have the same type of particles on the letf hand side moving both of them towards the Q-Ball.
Since we we have chosen the parameters to be the incident amplitudes we can choose that there is no incident
anti-particle, $C_2^L=0$. Using the matrix relations we find,
\begin{eqnarray}
C_1^R=\frac{\Exp{-2ik_\ep l}-\alpha_\ep^2\Exp{2ik_\ep l}}{1-\alpha_\ep^2}=T_{pp} \nonumber, \\
C_2^R=\frac{\alpha_\ep\Exp{-2ik_\ep l}-\alpha_\ep\Exp{2ik_\ep l}}{1-\alpha_\ep^2}=T_{p\bar{p}} \nonumber,
\end{eqnarray}
The transformation amplitude we are looking for is the $C_2^R$ coefficient. As long as $k_\ep$ is complex (see chap three), we
can easily see that this amplitude decreases as the Q-Ball's size increases. Here we considered the range where $\ep\geq\frac{\om_0}{2}$ and the 
solution is,
\begin{eqnarray}
\Psi_L=\int_{\frac{\om_0}{2}}^{\infty}d\ep\lp\ba{c}A(\ep)\Exp{-i(\ep+\frac{\om_0}{2})t}\Exp{i(\ep+\frac{\om_0}{2})t}\\0\ea\rp \nonumber \\
\Psi_R=\int_{\frac{\om_0}{2}}^{\infty}d\ep\lp\ba{c}A(\ep)T_{pp}\Exp{-i(\ep+\frac{\om_0}{2})t}\Exp{i(\ep+\frac{\om_0}{2})t}\\
A^\star(\ep)T_{p\bar{p}}^\star\Exp{i(\ep-\frac{\om_0}{2})t}\Exp{i(\ep-\frac{\om_0}{2})t}\ea\rp, \nonumber
\end{eqnarray}
where the $L$ and $R$ subscripts indicate on which side of the Q-Ball we are. At this point the $A$ expansion factor is considered to be
contiunous so it can be upgraded to an operator later on.
It is not the only range where transformation occurs we also have the other range $\ep\leq-\frac{\om_0}{2}$, with this time incident
particles coming from the right hand side of Q-Ball, we have
\begin{eqnarray}
\lp\ba{c} C_1^R \\ C_2^R \ea\rp=\frac{1}{1-\alpha^2}\lp\ba{cc} \Exp{-2ik_\ep l}-\alpha_\ep^2 \Exp{2ik_\ep l} & 
-\alpha_\ep\Exp{-2ik_\ep l}+\alpha_\ep\Exp{2ik_\ep l} \\\alpha_\ep\Exp{-2ik_\ep l}-\alpha_\ep\Exp{2ik_\ep l} & 
\Exp{2ik_\ep l}-\alpha_\ep^2 \Exp{-2ik_\ep l} \ea\rp\lp\ba{c} C_1^L \\ C_2^L \ea\rp. \nonumber
\end{eqnarray}
we have this time $C^1_R=0$, leading to the same solution for the transmisson amplitudes,
\begin{eqnarray}
\tilde{T}_{pp}=\frac{\Exp{-2ik_\ep l}-\alpha_\ep^2\Exp{2ik_\ep l}}{1-\alpha_\ep^2}=T_{pp} \nonumber, \\
\tilde{T}_{p\bar{p}}=\frac{\alpha_\ep\Exp{-2ik_\ep l}-\alpha_\ep\Exp{2ik_\ep l}}{1-\alpha_\ep^2}=T_{p\bar{p}} \nonumber,
\end{eqnarray}
even if the amplitudes are the same, wich is a effect of symmetry, we shall use tilde symbols to distinguish both contributions. In fact we would
like to keep in mind that this second contribution comes from waves having negative energy. These two contributions will mix together to build the
total solution. We have this time for solution,
\begin{eqnarray}
\Psi_L=\int^{-\frac{\om_0}{2}}_{-\infty}d\ep\lp\ba{c}B(\ep)\tilde{T}_{p\bar{p}}\Exp{-i(\ep+\frac{\om_0}{2})t}\Exp{i(\ep+\frac{\om_0}{2})z}\\
B^\star(\ep)\tilde{T}_{pp}^\star\Exp{i(\ep-\frac{\om_0}{2})t}\Exp{i(\ep-\frac{\om_0}{2})z}\ea\rp \nonumber \\
\Psi_R=\int^{-\frac{\om_0}{2}}_{-\infty}d\ep\lp\ba{c}0\\
B^\star(\ep)\Exp{i(\ep-\frac{\om_0}{2})t}\Exp{i(\ep-\frac{\om_0}{2})t}\ea\rp.
\end{eqnarray}
Let us take a look at the pieces on the left,
\begin{eqnarray}
\Psi_L&=&\int_{\frac{\om_0}{2}}^{\infty}d\ep u_1\Exp{-i(\ep+\frac{\om_0}{2})t}\Exp{i(\ep+\frac{\om_0}{2})z}A(\ep)+
\int^{-\frac{\om_0}{2}}_{-\infty}d\ep u_1\Exp{-i(\ep+\frac{\om_0}{2})t}\Exp{i(\ep+\frac{\om_0}{2})z}\tilde{T}_{p\bar{p}}B(\ep) \nonumber \\
&+&\int^{-\frac{\om_0}{2}}_{-\infty}d\ep u_2\Exp{i(\ep-\frac{\om_0}{2})t}\Exp{i(\ep-\frac{\om_0}{2})z}\tilde{T}_{pp}B^\star(\ep), \nonumber
\end{eqnarray}
where $u_1=\lp\ba{c}1\\0\ea\rp$ and $u_2=\lp\ba{c}0\\1\ea\rp$. We change the varible $\ep\rightarrow-\ep$ in the integrals on the
negative range to obtain,
\begin{eqnarray}
\Psi_L&=&\int_{\frac{\om_0}{2}}^{\infty}d\ep u_1\Exp{-i(\ep+\frac{\om_0}{2})t}\Exp{i(\ep+\frac{\om_0}{2})z}A(\ep)+
\int_{\frac{\om_0}{2}}^{\infty}d\ep u_1\Exp{i(\ep-\frac{\om_0}{2})t}\Exp{-i(\ep-\frac{\om_0}{2})z}\tilde{T}_{p\bar{p}}(-\ep)B(-\ep) \nonumber \\
&+&\int_{\frac{\om_0}{2}}^{\infty}d\ep u_2\Exp{-i(\ep+\frac{\om_0}{2})t}\Exp{-i(\ep+\frac{\om_0}{2})z}\tilde{T}_{pp}(-\ep)B^\star(-\ep). \nonumber
\end{eqnarray}
The next thing to do is change the variable $\ep+\frac{\om_0}{2}=\ep'$ and $\ep-\frac{\om_0}{2}=\ep'$ to obtain,
\begin{eqnarray}
\Psi_L&=&\int_{\om_0}^{\infty}d\ep u_1\Exp{-i\ep't}\Exp{i\ep'z}A(\ep'-\frac{\om_0}{2})+
\int_{0}^{\infty}d\ep u_1\Exp{i\ep't}\Exp{-i\ep'z}\tilde{T}_{p\bar{p}}(-\frac{\om_0}{2}-\ep)B(-\frac{\om_0}{2}-\ep) \nonumber \\
&+&\int_{\om_0}^{\infty}d\ep u_2\Exp{-i\ep't}\Exp{-i\ep'z}\tilde{T}_{pp}(-\ep'+\frac{\om_0}{2})B^\star(-\ep'+\frac{\om_0}{2}). \nonumber
\end{eqnarray}
Due to the energy shift comming from the Q ball we have two different bounds on the integration $\ep\in[\om_0,\infty[$ and
$\ep\in[0,\infty[$. These two ranges will be of great interest, since each one can be identified to a different particle. Before we 
construct the quantised solution we need to do the same transformations to the solution on right-hand side.
\begin{eqnarray}
\Psi_R&=&\int_{\frac{\om_0}{2}}^{\infty}d\ep u_1\Exp{-i(\ep+\frac{\om_0}{2})t}\Exp{i(\ep+\frac{\om_0}{2})z}T_{pp}A(\ep)+
\int^{\frac{\om_0}{2}}_{\infty}d\ep u_2\Exp{i(\ep-\frac{\om_0}{2})t}\Exp{i(\ep-\frac{\om_0}{2})z}T_{p\bar{p}}^\star A^\star(\ep) \nonumber \\
&+&\int^{-\frac{\om_0}{2}}_{-\infty}d\ep u_2\Exp{i(\ep-\frac{\om_0}{2})t}\Exp{i(\ep-\frac{\om_0}{2})z}B^\star(\ep), \nonumber
\end{eqnarray}
after the variable changes and the sign change we obtain,
\begin{eqnarray}
\Psi_R&=&\int_{\om_0}^{\infty}d\ep u_2\Exp{-i\ep't}\Exp{i\ep'z}B(-\ep'+\frac{\om_0}{2})+
\int_{0}^{\infty}d\ep u_2\Exp{i\ep't}\Exp{-i\ep'z}T_{p\bar{p}}^\star(\frac{\om_0}{2}+\ep)A(\frac{\om_0}{2}+\ep) \nonumber \\
&+&\int_{\om_0}^{\infty}d\ep u_1\Exp{-i\ep't}\Exp{-i\ep'z}T_{pp}(\ep'-\frac{\om_0}{2})A(\ep'-\frac{\om_0}{2}). \nonumber
\end{eqnarray}
We see here that the $B(-\ep'+\frac{\om_0}{2})$ and $A(\ep'-\frac{\om_0}{2})$ are in the same integration range and therefor will
corespond to the same particles.
We now have both solutions one on each side of the Q-Ball, the last step of construction will be quantization. Quantization will be done
identifying the parts coresponding to particles and the parts coresponding to anti-particles. It is easy to identify the particles
and the anti-particles since the integration domain is well known and definied we have,
\begin{eqnarray}
\Psi_L&=&\int_{\om_0}^{\infty}d\ep\Exp{-i\ep t}\{\Exp{i\ep z}u_1+\Exp{-i\ep z}\tilde{T}_{pp}(\frac{\om_0}{2}-\ep)u_2\}a_l(\ep-\frac{\om_0}{2}) \nonumber \\
&+&\int_{0}^{\infty}d\ep\Exp{i\ep t}\{\Exp{-i\ep z}\tilde{T}_{p\bar{p}}(\frac{\om_0}{2}+\ep)u_1\}b_l^\dagger(\ep+\frac{\om_0}{2}) \nonumber
\end{eqnarray}
for the solution on the left and
\begin{eqnarray}
\Psi_R&=&\int_{\om_0}^{\infty}d\ep\Exp{-i\ep t}\{\Exp{-i\ep z}u_2a_r(\frac{\om_0}{2}-\ep)+\Exp{i\ep z}T_{pp}(\frac{\om_0}{2}-\ep)u_1
a_r^\dagger(\frac{\om_0}{2}-\ep)\} \nonumber \\
&+&\int_{0}^{\infty}d\ep\Exp{i\ep t}\{\Exp{i\ep z}T_{p\bar{p}}(\frac{\om_0}{2}+\ep)u_1\}b_r^\dagger(\ep+\frac{\om_0}{2}), \nonumber
\end{eqnarray}
for the solution on the right hand side of the Q-Ball. During the construction we used,
\begin{eqnarray}
B(k)=a(k)\theta(k)+b^\dagger(k)\theta(-k). \nonumber
\end{eqnarray}
We keep for the moment the $l$ and $r$ subscript on the operators to identify the side of the Q ball they correspond. During this construction we did not 
normalise any solution. The normalisation of solutions is a simple task \cite{clark}, and will not be needed in the rest of the discussion. We just need
to keep in mind that all the waves packets are normalised in the way that the sum of the reflected amplitude and the transmitted one is equal to one. 

Now that we have the total solution in terms of creation and anihilisation operators we can compute anything we want, by applying the
proper operator valued observable. We can see that the energy of the transmited wave is the same as the energy of the incoming wave, while the 
energy of the reflected anti-particle is shifted by a factor of $\om_0$. This shift in energy, one component shifted downwards and the
other one shifted upwards is the reason why we found finite evaporation ranges. This property is due to the Yukawa coupling inside
the Q-Ball. We can choose a state where there is no incoming particles on the right so our transmission coeficients can be identified
to reflection ones. In fact this construction can be used in the full energy range. The difficulty will be to seperate the phenomenons linked
to diffusion and those linked to evaporation.
\subsection{Diffusion on a Q ball}
Now that we have the quantised solutions we can try to study diffusion on a Q ball. We need to use the total solution having
no incident wave on the right-hand side of the Q ball. We write,
\begin{eqnarray}
\Psi_L&=&\int_{\om_0}^{\infty}d\ep\Exp{-i\ep t}\{\Exp{i\ep z}u_1+\Exp{-i\ep z}\tilde{T}_{pp}(\frac{\om_0}{2}-\ep)u_2\}a_l(\ep-\frac{\om_0}{2}) \nonumber \\
&+&\int_{0}^{\infty}d\ep\Exp{i\ep t}\{\Exp{-i\ep z}\tilde{T}_{p\bar{p}}(\frac{\om_0}{2}+\ep)u_1\}b_l^\dagger(\ep+\frac{\om_0}{2}) \nonumber \\
&+&\int_{\om_0}^{\infty}d\ep\Exp{-i\ep t}\{\Exp{i\ep z}T_{pp}(\frac{\om_0}{2}-\ep)u_1a_r^\dagger(\frac{\om_0}{2}-\ep)\} \nonumber \\
&+&\int_{0}^{\infty}d\ep\Exp{i\ep t}\{\Exp{i\ep z}T_{p\bar{p}}(\frac{\om_0}{2}+\ep)u_1\}b_r^\dagger(\ep+\frac{\om_0}{2}), \nonumber
\end{eqnarray}
the construction of the Bogolubov transformation linking the far past, the incident wave, to the far future reads :
\begin{eqnarray}
c_{out}&=&T_{pp}(\frac{\om_0}{2}-\ep)a_l(\ep-\frac{\om_0}{2})+T_{p\bar{p}}(\frac{\om_0}{2}+\ep)b_l^\dagger(\ep+\frac{\om_0}{2}) \nonumber \\
&+&T_{pp}(\frac{\om_0}{2}-\ep)a_r^\dagger(\frac{\om_0}{2}-\ep)+T_{p\bar{p}}(\frac{\om_0}{2}+\ep)b_r^\dagger(\ep+\frac{\om_0}{2})\nonumber.
\end{eqnarray}
We here distinguish two energy ranges, if $\ep\geq\frac{\om_0}{2}$ the $a_r^\dagger(\frac{\om_0}{2}-\ep)$ operator becomes $a_r(\ep-\frac{\om_0}{2})$, while if
$\ep\leq\frac{\om_0}{2}$ it is the $a_l(\ep-\frac{\om_0}{2})$ becomming $a_l^\dagger(\frac{\om_0}{2}-\ep)$. As we said before we had to separate both regimes
in order to see what type of diffusion we have.

First we shall consider an infinite Q ball, so the right-hand side does not exist, and we have
\begin{eqnarray}
c_{out}&=&T_{pp}(\frac{\om_0}{2}-\ep)a_l(\ep-\frac{\om_0}{2})+T_{p\bar{p}}(\frac{\om_0}{2}+\ep)b_l^\dagger(\ep+\frac{\om_0}{2}), \nonumber \\
\end{eqnarray}
when we constructed the solution the incident particle had energy bigger than $\om_0$. So in this range $\ep-\frac{\om_0}{2}$ is allways bigger than zero,
and the $a$ operator remains an anihilation operator. We have now in the final state (in the far future),
\begin{eqnarray}
_{out}<0|c^\dagger c|0>_{out}=|T_{p\bar{p}}|^2<\bar{p}|\bar{p}>.
\end{eqnarray}
This simply is a particle being reflected into an anti-particle to conserve helicity. To study the rest of the possibilities we shall use a small
sized Qball. 

The first possibility stands for $\ep\geq\om_0/2$ we need to change the type of the second $a$ operator, $a_r^\dagger(\frac{\om_0}{2}-\ep)$ becomes
$a_r(\ep-\frac{\om_0}{2})$ :
\begin{eqnarray}
c_{out}&=&T_{pp}(\frac{\om_0}{2}-\ep)a_l(\ep-\frac{\om_0}{2})+T_{p\bar{p}}(\frac{\om_0}{2}+\ep)b_l^\dagger(\ep+\frac{\om_0}{2}) \nonumber \\
&+&T_{pp}(\frac{\om_0}{2}-\ep)a_r(\ep-\frac{\om_0}{2})+T_{p\bar{p}}(\frac{\om_0}{2}+\ep)b_r^\dagger(\ep+\frac{\om_0}{2})\nonumber.
\end{eqnarray}
Applying this to the vaccum state in the far future we have,
\begin{eqnarray}
_{out}<0|c^\dagger c|0>_{out}=|\tilde{T}_{p\bar{p}}|^2{_l<\bar{p}|\bar{p}>_l}+|T_{p\bar{p}}|^2{_r<\bar{p}|\bar{p}>_r}.
\end{eqnarray}
\begin{figure}[h]
\begin{center}
\input{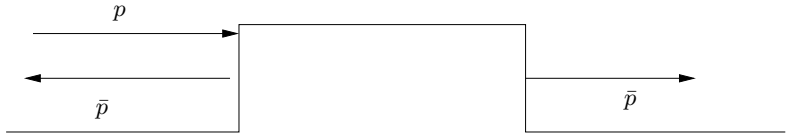}
\caption{Diffusion on a Q ball for an incident particle having energy bigger than $\om_0$.}
\label{diff1}
\end{center}
\end{figure}

The second possibility will be for a particle having energy lower than $\om_0/2$ we have to change the type of the first $a$, $a_l(\ep-\frac{\om_0}{2})$
becomes $a_l^\dagger(\frac{\om_0}{2}-\ep)$. We write :
\begin{eqnarray}
c_{out}&=&T_{pp}(\frac{\om_0}{2}-\ep)a_l^\dagger(\frac{\om_0}{2}-\ep)+T_{p\bar{p}}(\frac{\om_0}{2}+\ep)b_l^\dagger(\ep+\frac{\om_0}{2}) \nonumber \\
&+&T_{pp}(\frac{\om_0}{2}-\ep)a_r^\dagger(\frac{\om_0}{2}-\ep)+T_{p\bar{p}}(\frac{\om_0}{2}+\ep)b_r^\dagger(\ep+\frac{\om_0}{2})\nonumber.
\end{eqnarray}
This time we have four particles in the final state, it is the combination of evaporation and diffusion.
\begin{figure}[h]
\begin{center}
\input{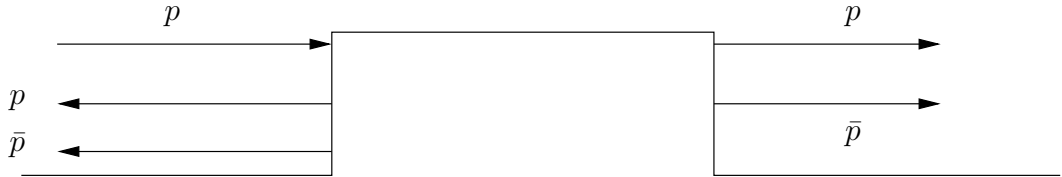}
\caption{Diffusion on a Q ball for an incident particle having energy smaller than $\om_0/2$.}
\label{diff2}
\end{center}
\end{figure}

The third possiblity would be for incident fermions having their energy in the range $\ep\in[\om_0/2;\om_0]$, we shall not study this range since we
only considered incident fermions with enrgy bigger than $\om_0$ or smaller than $\om_0/2$. But this range does not introduce any new or difficult
construction. These construction can also be used to compute evaporation rate we just need to separate both constructions and then identify wich
contribution is diffusion and wich one is evaporation. 


\subsection{Diffusion on a Q ball massive case.}
Starting with the solution for a Q ball interacting with a fermion we have,
{\small
\begin{eqnarray}
\Psi&=&\int_{M_D+\frac{\om_0}{2}}^\infty d\ep
\{\Exp{i(\ep-\frac{\om_0}{2})t}[\Exp{i\bar{p}_1x}u_{\bar{p_1}}^{up}+r_{11}\Exp{-i\bar{p}_1x}u_{-\bar{p_1}}^{up}]b^\dagger_{\bar{p}_1}  
+\Exp{-i(\ep+\frac{\om_0}{2})t}[r_{12}^\star\Exp{i\bar{p}_2^\star x}(u_{-\bar{p_2}}^{down})^\star]b_{\bar{p}_1}\}, \nonumber\\
&+&\int_{M_D-\frac{\om_0}{2}}^\infty d\ep\{\Exp{i(\ep-\frac{\om_0}{2})t}[r_{21}\Exp{-i\bar{p}_1x}u_{-\bar{p_1}}^{up}]a^\dagger_{\bar{p}_2}
+\Exp{-i(\ep+\frac{\om_0}{2})t}[\Exp{-i\bar{p}_2^\star x}u_{\bar{p_2}}^\star+r_{22}^\star\Exp{i\bar{p}_2^\star x}(u_{-\bar{p_2}}^{down})^\star]
a_{\bar{p}_2}\}, 
\nonumber\\
&+&\int_{M_D+\frac{\om_0}{2}}^\infty d\ep\{\Exp{i(\ep-\frac{\om_0}{2})t}[\tilde{t}_{11}\Exp{-i\bar{p}_1x}u_{-\bar{p_1}}^{up}]b^\dagger_{-\bar{p}_1} 
+\Exp{-i(\ep+\frac{\om_0}{2})t}[\tilde{t}^\star_{12}\Exp{i\bar{p}_2^\star x}(u_{-\bar{p_2}}^{down})^\star]b_{-\bar{p}_1}\}, \nonumber\\
&+&\int_{M_D-\frac{\om_0}{2}}^\infty d\ep\{\Exp{i(\ep-\frac{\om_0}{2})t}[\tilde{t}_{21}\Exp{-i\bar{p}_1x}u_{-\bar{p_1}}^{up}]a^\dagger_{-\bar{p}_2} 
+\Exp{-i(\ep+\frac{\om_0}{2})t}[\tilde{t}^\star_{22}\Exp{i\bar{p}_2^\star x}(u_{-\bar{p_2}}^{down})^\star]a_{-\bar{p}_2}\},\nonumber
\end{eqnarray}
}
as before this solution has two different integration ranges one for each type of particles. We shall now only consider the range where
$\ep\geq M_D+\frac{\om_0}{2}$ in this range we only have diffusion and to not need to study the interaction of both, diffusion and evaporation.
We have ,
{\small
\begin{eqnarray}
\Psi&=&\int_{M_D+\frac{\om_0}{2}}^\infty d\ep
\{\Exp{i(\ep-\frac{\om_0}{2})t}[\Exp{i\bar{p}_1x}u_{\bar{p_1}}^{up}+r_{11}\Exp{-i\bar{p}_1x}u_{-\bar{p_1}}^{up}]b^\dagger_{\bar{p}_1}  
+\Exp{-i(\ep+\frac{\om_0}{2})t}[r_{12}^\star\Exp{i\bar{p}_2^\star x}(u_{-\bar{p_2}}^{down})^\star]b_{\bar{p}_1}\}, \nonumber\\
&+&\int_{M_D+\frac{\om_0}{2}}^\infty d\ep\{\Exp{i(\ep-\frac{\om_0}{2})t}[r_{21}\Exp{-i\bar{p}_1x}u_{-\bar{p_1}}^{up}]a^\dagger_{\bar{p}_2}
+\Exp{-i(\ep+\frac{\om_0}{2})t}[\Exp{-i\bar{p}_2^\star x}u_{\bar{p_2}}^\star+r_{22}^\star\Exp{i\bar{p}_2^\star x}(u_{-\bar{p_2}}^{down})^\star]
a_{\bar{p}_2}\}, 
\nonumber
\end{eqnarray}
}
the only reason why we used a very big Q ball is that we have less terms to deal with. With this solution we can as in the previous chapter construct
a bogolubov transformation considering that in the far future only out going waves survies. We have this time :
\begin{eqnarray}
r_{11}b^\dagger_{\bar{p}_1}+r_{12}^\star b_{\bar{p}_1}+r_{21}a^\dagger_{\bar{p}_2}+r_{22}^\star a_{\bar{p}_2}=c_{out}.
\end{eqnarray}
If we want $c_{out}$ to have the same anti-commutation relations as $a$ and $b$ we need to normalise the operators with :
\begin{eqnarray}
a=\frac{a}{|r_{22}|^2+|r_{21}|^2}, \\
b=\frac{b}{|r_{12}|^2+|r_{11}|^2},
\end{eqnarray}
There is no obvious  reasons why $|r_{12}|^2=|r_{21}|^2$ and $|r_{11}|^2=|r_{22}|^2$. These equalities will depend on the type of Q ball we study, they depend on
the way the Q ball mixes the particles and anti-particle energies.
We can use this $c$ operator to compute the number of particles in the final state when the original one was the vacuum state,
\begin{eqnarray}
_{out}<0|c^\dagger c|0>_{out}&=&|r_{21}|^2_{in}<0|aa^\dagger|0>_{in}+|r_{11}|^2_{in}<0|bb^\dagger|0>_{in}, \nonumber \\
&=&|r_{21}|^2<p|p>+|r_{11}|^2<ap|ap>.
\end{eqnarray}
We have this time, when fermions are massive, a particle and an anti-particle in the final state. Those two reflected particules will be part of the state
combining evaporation and diffusion. This combination of evaporation and diffusion is more complicated to study since we need to use the full solution and
be carefull with the energy ranges and the type of operator linked to them.
\section{Conclusions.}
As expected the interaction of Q balls and fermions can be separated into two different cases. The first one stands for interaction between the Q ball and fermions 
having their energy lying outside the evaporation rate. In this case we demonstrated that the interaction of Q balls and matter is nothing but standard diffusion.
If we have an incident fermion we have a reflected and transmitted anti-fermion. The transmitted particles has its energy shifted by a $\om_0/2$ factor due to 
interaction with the Q ball.

The second case happens when the fermion interacting with the Q ball has its energy lying inside the evaporation range. In this case we have a superposition of
two phenomenoms the first one is diffusion while the second one is evaporation. It seems that both phenomenoms to not interfear together and be considered separately.
This fact is important in the way that it provides a new approach to compute Q ball evaporation rates. This new approach would be to study the diffusion of a particle
on the Q ball's surface and find the two ranges where we have diffusion and both diffusion and evaporation. Then we just need to substract both amplitudes to obtain
evaporation range.

The interaction with massive fermions does not introduce any new facts. The calculations becomes more complicated but the separation into two ranges stands the same
way, exept this time the evaporation range is different since it depends on the fermion mass. In this case the other difference comes from the fact that we do not
use helicity conservation to find the reflected waves.

\end{document}